\documentclass[aps,pra,twocolumn,superscriptaddress, notitlepage,10pts, showkeys]{revtex4-1}
\usepackage{graphicx} 
\usepackage{lipsum}
\usepackage{orcidlink}
\usepackage{xcolor}
\usepackage{subcaption}
\usepackage{url}
\usepackage{booktabs}  
\usepackage{tabularx}   
\usepackage[table]{xcolor}
\usepackage{amssymb}

\begin{document}

\title{Terrestrial campaign towards space-to-ground quantum communications with a space-qualified entangled photon-pair system}

\author{Gianluca De Santis \orcidlink{0009-0005-2393-6076}}
\email{gianluca.desantis@tii.ae}
\affiliation{Quantum Research Center, Technology Innovation Institute, PO Box 9639 Abu Dhabi, United Arab Emirates}
\affiliation{Sorbonne Université, CNRS, LIP6, 4 Place Jussieu, Paris F-75005, France}

\author{Jia Boon Chin \orcidlink{0009-0002-5468-413X}}
\affiliation{Center for Quantum Technologies, National University of Singapore, Singapore 117543, Singapore}

\author{Srihari Sivasankaran \orcidlink{0000-0002-4186-3115}}
\affiliation{SpeQtral Pte., Ltd., Singapore}

\author{Konstantin Kravtsov \orcidlink{0000-0003-4499-4089}}
\affiliation{Quantum Research Center, Technology Innovation Institute, PO Box 9639 Abu Dhabi, United Arab Emirates}

\author{Chin Chean Lim}
\affiliation{SpeQtral Pte., Ltd., Singapore}

\author{Aitor Villar \orcidlink{0000-0001-8248-6830}}
\affiliation{SpeQtral Pte., Ltd., Singapore}

\author{Robert Bedington \orcidlink{0000-0001-8033-8244}}
\affiliation{SpeQtral Pte., Ltd., Singapore}

\author{Sana Amairi-Pyka \orcidlink{0000-0001-9502-4176}}
\affiliation{Quantum Research Center, Technology Innovation Institute, PO Box 9639 Abu Dhabi, United Arab Emirates}

\author{Eleni Diamanti \orcidlink{0000-0003-1795-5711}}
\affiliation{Sorbonne Université, CNRS, LIP6, 4 Place Jussieu, Paris F-75005, France}

\author{Alexander Ling \orcidlink{0000-0001-5866-1141}}
\affiliation{Center for Quantum Technologies, National University of Singapore, Singapore 117543, Singapore}
\affiliation{Department of Physics, National University of Singapore, Singapore 117551, Singapore}

\author{James A. Grieve \orcidlink{0000-0002-2800-8317}}
\affiliation{Quantum Research Center, Technology Innovation Institute, PO Box 9639 Abu Dhabi, United Arab Emirates}

\begin{abstract}

Realizing a global quantum internet relies on the deployment of robust satellite-based entanglement distribution links. While pioneering demonstrations have established the feasibility of such links, the transition to operational infrastructure demands the validation of robust, integrated space-to-ground architectures. Here, we report on a free-space Quantum Key Distribution experiment conducted over a 1.8~km free-space link using an engineering model of the quantum payload onboard the SpeQtre satellite and the Abu Dhabi Quantum Optical Ground Station. By implementing a BBM92 protocol with polarization-entangled photons, a mean secret key rate of 7.57 kbps with a mean quantum bit error rate of $4.78\,\% \pm 0.24 \, \%$ was produced. The deployed system featured spectral and spatial filtering approaches identical to those in the space segment, demonstrating their operation in a terrestrial nighttime free-space channel. These results confirm the compatibility between the ground and space segments, establishing a critical performance baseline for the SpeQtre mission and future space-based, large-scale quantum networks.
\section*{Key Words}
Quantum Key Distribution, Satellite quantum communication, Free-space quantum communications, Entanglement distribution,  Optical Ground Station. 
\end{abstract}

\maketitle

\section{Introduction}

The distribution of quantum resources is fundamental to the development of future quantum networking infrastructure \cite{wei2022towards}. Various pathways have been proposed to realize a quantum internet \cite{kimble2008quantum, simon2017towards, wehner2018quantum}, wherein diverse physical architectures will coexist to implement novel tasks in computation, communication, and sensing. Quantum networking constitutes the connective framework for these heterogeneous devices, with photons acting as the optimal carriers for such interactions.

Beyond simple state transfer, the distribution of entangled states enables a broader class of advanced protocols, such as blind quantum computing \cite{broadbent2009universal}, distributed quantum computing \cite{yimsiriwattana2004distributed}, and distributed quantum sensing \cite{zhang2021distributed}. These capabilities extend the utility of the network far beyond secure communication, laying the groundwork for a truly interconnected quantum ecosystem. In the field of communication, a primary application of quantum resource distribution is Quantum Key Distribution (QKD), which allows for the establishment of information-theoretic secure keys between two parties \cite{gisin2002quantum}. While QKD represents only one of many applications enabled by quantum networking, it serves as a critical benchmark for demonstrating the maturity of quantum technologies.

Globally, several QKD networks have demonstrated the capability to distribute quantum resources. Fiber-based prepare-and-measure implementations have shown remarkable performance in terms of distance \cite{yin2016measurement,liu2023experimental}, user capacity \cite{frohlich2015quantum}, and key-rate generation \cite{yuan201810, li2023high}. Conversely, entanglement-based solutions offer enhanced security by providing a pathway toward device-independent protocols \cite{zapatero2023advances} and increased resilience against eavesdropping strategies \cite{acin2006bell}. Furthermore, entanglement distribution is a prerequisite for long-distance communication via quantum repeater protocols \cite{pan1998experimental,azuma2023quantum}.

However, realizing a global quantum network necessitates the distribution of resources across intercontinental distances \cite{de2023satellite}. For such scales, optical fiber is limited by fundamental attenuation \cite{pirandola2017fundamental} as well as geopolitical and infrastructural constraints \cite{oecd_rights_of_way_fibre_2008, au_crossborder_fibre_permits}. Consequently, satellite-based links emerge as a viable solution for achieving global connectivity \cite{de2023satellite}. Space-based QKD has undergone extensive development over the last two decades \cite{bedington2017progress, sidhu2021advances}. Current orbital demonstrations predominantly utilize prepare-and-measure architectures \cite{liao2017satellite, liao2017space, liao2018satellite, li2022space, khmelev2024eurasian, li2025microsatellite}, which have successfully established intercontinental links and achieved secure key generation rates on the order of kbps.

Conversely, entanglement-based (EB) protocols remain at a more nascent stage of orbital deployment and face significant implementation hurdles. Initial efforts focused on validating the feasibility of orbital entanglement distribution via Bell inequality violations \cite{yin2017satellite,villar2020entanglement}, while subsequent single-downlink and double-downlink QKD implementations yielded secret key rates in the bps range \cite{yin2017satellite_b, yin2020entanglement}. In single-downlink configurations, the satellite locally measures one photon from an entangled pair while transmitting the other photon to a ground station, whereas double-downlink schemes simultaneously distribute the entangled photons to two separate ground stations. Notably, the double-downlink demonstration across a 1120~km baseline reported a QBER of $4.5\% \pm 0.4\%$ with an asymptotic key rate of 0.43~bps (0.12~bps under finite-size effects) \cite{yin2020entanglement}.

\begin{figure}
    \centering
    \includegraphics[width=0.99\linewidth]{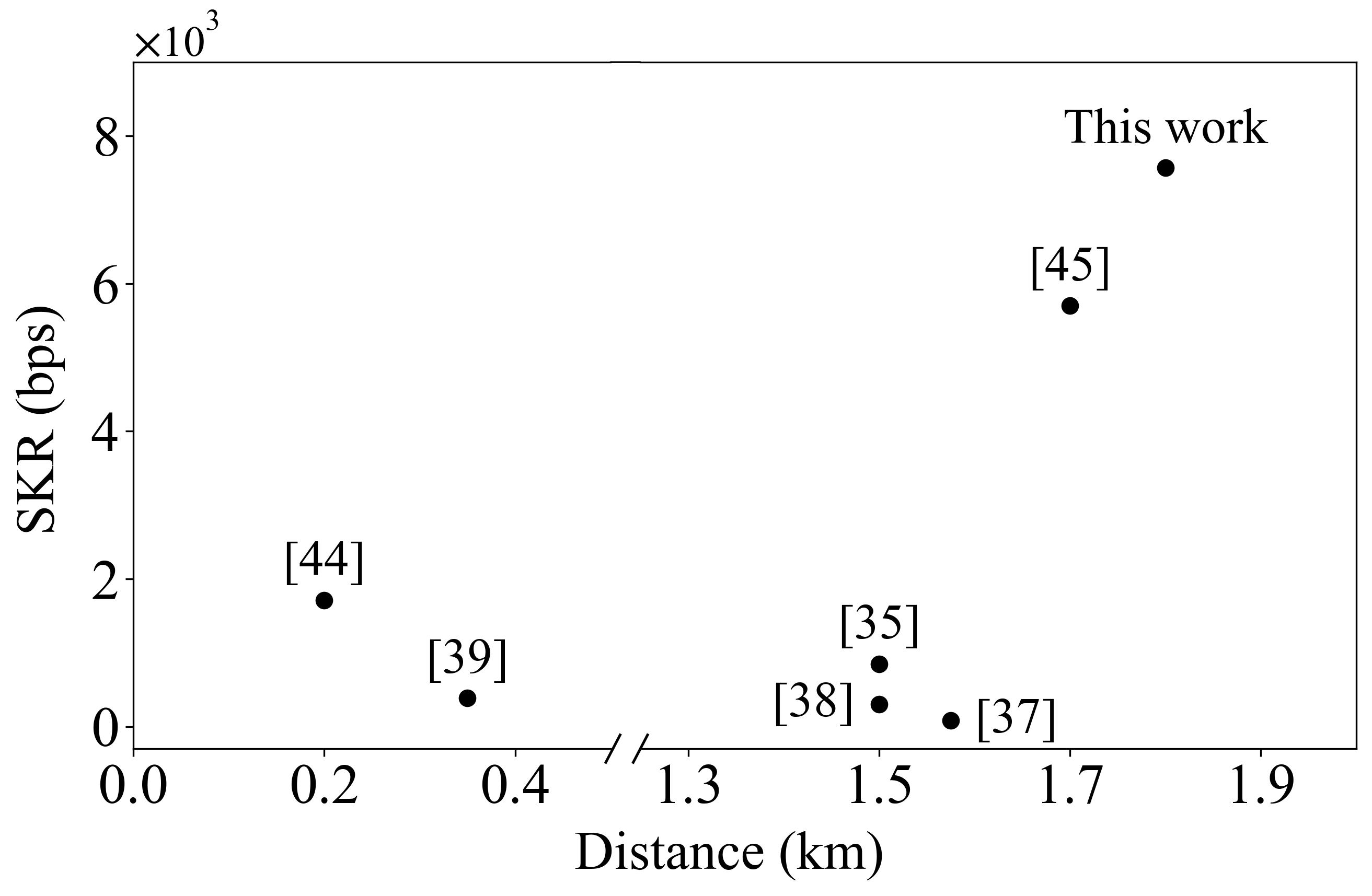}
    \caption{Comparison of Secret Key Rate (SKR) versus distance for free-space polarization-entangled QKD systems over links up to 2 km. Data points represent reported results from the existing literature alongside the current study. The SKR values for both ``This work'' and Ref. [45] are calculated utilizing finite-size analysis with a 5-minute aggregation time.}
    \label{fig:lit_rev}
\end{figure}

\begin{figure*}
    \centering
    \includegraphics[width=0.70\linewidth]{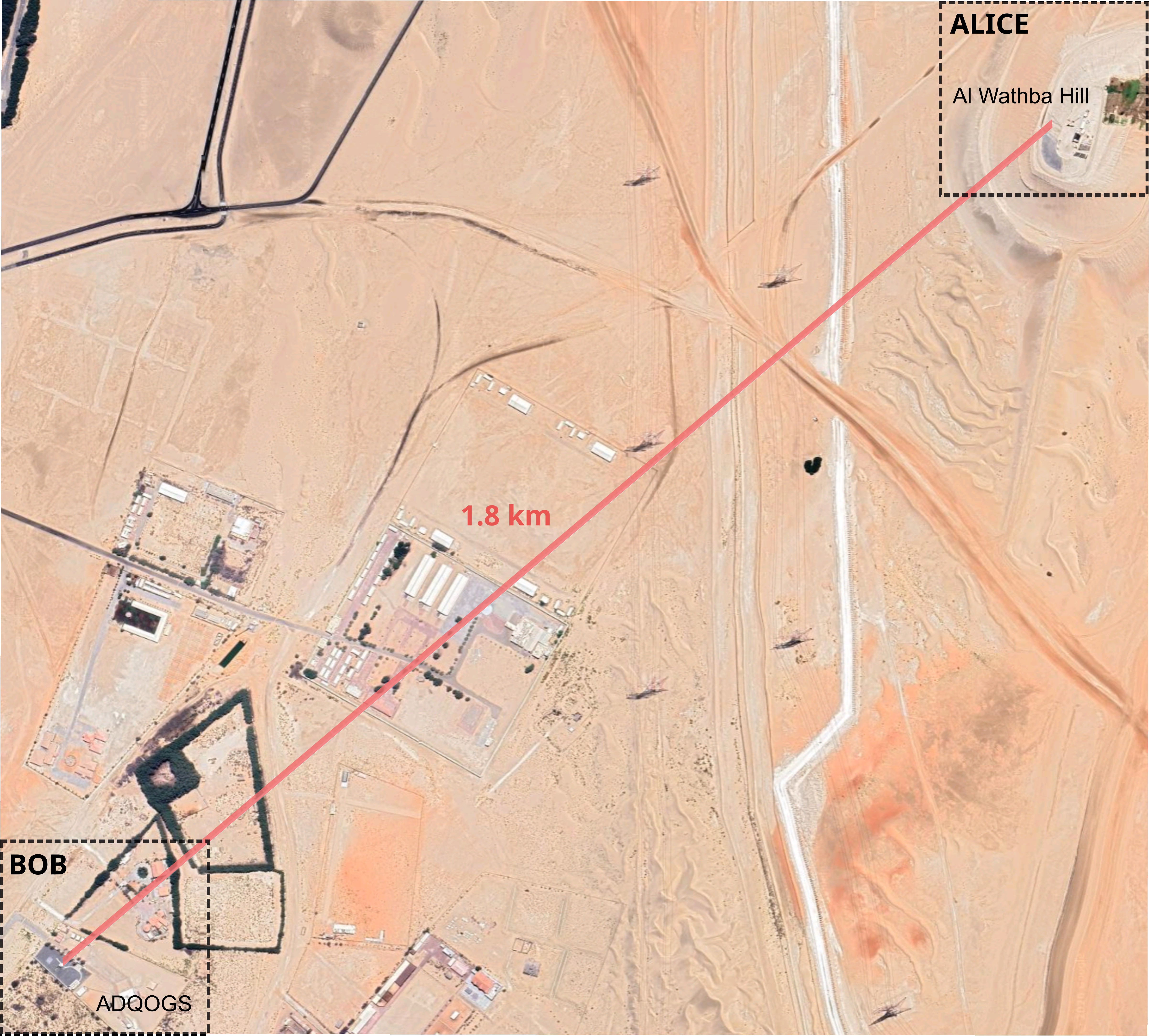}
    \caption{Satellite image of the link between Al Wathba hill (Alice) and ADQOGS (Bob) in Abu Dhabi. Image from Google Maps.}
    \label{fig:map}
\end{figure*}

Terrestrial horizontal links have served as essential testbeds for maturing the technology. Early experimental efforts successfully verified the viability of the distribution of polarization-entangled photons over a free-space channel under various atmospheric regimes \cite{peng2005experimental, marcikic2006free, ursin2007entanglement, erven2008entangled, ling2008experimental, peloso2009daylight, scheidl2009feasibility, erven2012studying, cao2013entanglement}. More recently, research has progressed toward optimizing link performance and environmental resilience. Recent studies have demonstrated strategies for achieving high key rates using adaptive source parameters \cite{ecker2021strategies}, quantified the impact of atmospheric aerosols on signal extinction \cite{mishra2022bbm92}, and showcased the feasibility of metropolitan-scale free-space networks using deployable, portable terminals \cite{krvzivc2023towards}.
While these studies have significantly advanced the state of the art, terrestrial demonstrations conducted to date have largely utilized specialized or ad hoc experimental configurations tailored for terrestrial flexibility and specific proof-of-concept objectives \cite{Rozenman2026, Trinh2018, Mehic2020}. As the field matures toward large-scale network integration, there is a critical need to transition these bespoke setups into standardized hardware architectures.
Bridging the gap between proof-of-concept demonstrations and operational satellite networks requires the rigorous characterization of quantum hardware under realistic field conditions. Specifically, validating the performance of state-of-the-art Optical Ground Stations (OGSs) and quantum payloads is crucial for establishing the reliability of entanglement-based links in space. 

In this work, we demonstrate a free-space QKD experiment based on the BBM92 protocol over a 1.8~km horizontal link during nighttime operations. The transmitter node utilizes the quantum light source and receiver from an engineering model of the SpeQtre satellite quantum payload \cite{villar2026speqtre}, emulating a single-downlink architecture, while the receiver node employs the Abu Dhabi Quantum Optical Ground Station (ADQOGS). By implementing spatial and spectral filtering identical to that of the full-scale satellite-to-ground mission, we obtained a secret key rate (SKR) in the kbps range. The observed SKR was the highest among free-space polarization-entangled QKD demonstrations up to 2~km (Fig.~\ref{fig:lit_rev}). A direct quantitative comparison of the reported SKRs is complicated by differences in channel loss, system architecture, post-processing procedures, and experimental conditions. Hence, we regard this comparison primarily as a performance benchmark rather than as a rigorous quantitative ranking of the reported experimental results. The measured terrestrial performance also provides a basis for assessing the expected operation of the same architecture over a space-to-ground link. We therefore extrapolate the results to a Low Earth Orbit (LEO) link scenario. This experimental configuration closely mimics the architecture intended for space-based quantum networking, providing critical insights for the deployment of global quantum communication networks.

\section{Results}

\subparagraph{Deployment}
A horizontal free-space QKD experiment was conducted at night under moderate atmospheric conditions on the 21st November 2025. The line-of-sight link spanned 1.8~km, with the transmitter (Alice) situated at an altitude of approximately 175~m above sea level (asl) and the receiver (Bob) positioned at approximately 50~m asl. The link was established in a semi-urban desert environment approximately 40~km from the city center of Abu Dhabi, United Arab Emirates (Fig.~\ref{fig:map}). This location provides an optimal trade-off between manageable background optical noise and logistical proximity to the city \cite{10.52202/083082-0038}. Alice and Bob were deployed as static, fixed outdoor nodes.

\subparagraph{Space-qualified quantum light source and receiver}
Alice’s node, located on the Al Wathba hill, comprised a source of polarization-entangled photon pairs and a quantum receiver for polarization analysis. These are space-qualified engineering models of the SpeQtre satellite quantum payload. 

The source is based on Type-0 spontaneous parametric down-conversion (SPDC) and utilizes a linear displacement interferometer design~\cite{Lohrmann2020, villar2026speqtre}. Photon pairs at wavelengths 780~nm (signal) and 842~nm (idler) with 5~nm FWHM spectrum are produced with a target state of $|\Phi^-\rangle=\frac{1}{\sqrt{2}}(|HH\rangle-|VV\rangle)$. Signal and idler photons are split via a fiber-based wavelength division multiplexer. The idler photons were locally projected into two mutually unbiased bases and detected using four single-photon silicon detectors in the quantum receiver. The signal photons were routed to a portable optical telescope with an aperture of 135~mm for beam transmission. The resulting $1/e^2$ optical spot size at the receiver plane at ADQOGS was approximately 100~mm in diameter.

\subparagraph{ADQOGS}
Bob’s node was housed within the ADQOGS, a versatile, fully automated facility designed for optical and quantum communications \cite{amairi2024versatile}. It incorporates an 80~cm Ritchey–Chrétien telescope equipped with a Quantum Acquisition and Tracking System (QATS). Photons collected by the primary mirror are routed into the backend optics, where a dichroic mirror separates the quantum channel from the classical beacon. The Quantum Module (QM) features four single-photon silicon detectors (SPDs) configured to measure the polarization in two mutually unbiased polarization bases. To mitigate background noise, the module employs a 16.6~nm FWHM spectral filter centered at 785~nm. Additionally, the field of view is restricted to $0.0028^\circ$ (10 arcsec in the sky) to implement spatial filtering. Under these conditions, the measured background per channel ranges from 150 to 1500~counts per second during new-moon and full-moon nights, respectively \cite{10.52202/083082-0038}.
The station is also equipped with a GPS-synchronized rubidium master clock and a weather station for atmospheric monitoring.

\begin{figure*}
    \centering
    \includegraphics[width=0.9\linewidth]{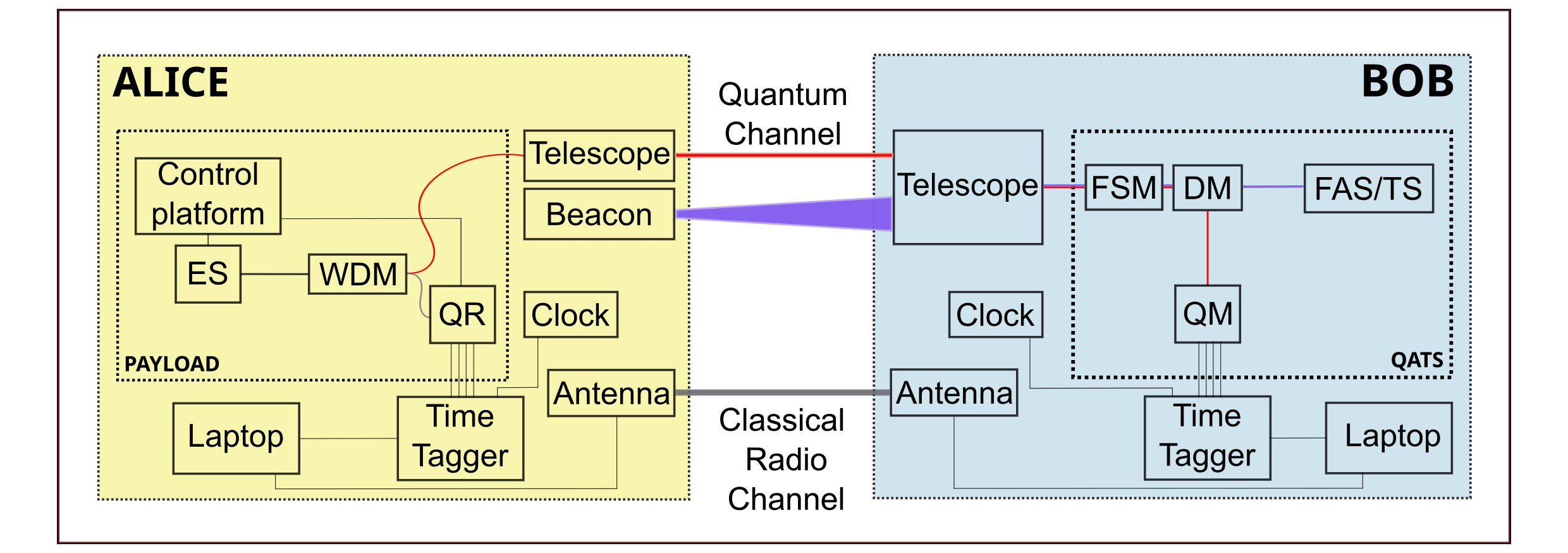}
    \caption{Scheme of the deployed setup. ES: Polarization-entangled photon-pair source. WDM: Wavelength division multiplexer. QR: Quantum receiver. FSM: Fast Steering Mirror. DM: Dichroic mirror. QM: Quantum module. FAS/TS: Fine acquisition and tracking sensor.}
    \label{fig:setup}
\end{figure*}

\subparagraph{QKD Link}
Using the setup described in Fig.~\ref{fig:setup}, an entanglement-based BBM92 QKD protocol was implemented.
Synchronization between the two nodes was achieved by calculating the cross-correlation function of the timestamps on each Time Tagger, which were disciplined using precision clock sources. Classical bi-directional communication was enabled by a pair of long-range radio bridge antennas. 
The initial pointing of Alice's telescope was achieved using a 780~nm alignment laser beam transmitted through the SPDC optical path. Following this, the alignment beam was replaced with the quantum signal from the photon-pair source, and the signal throughput was maximized by manual fine-tuning at both sides. The quantum bit error rate (QBER) was minimized by adjusting the polarization controllers inside the quantum payload. 
Throughout the experiment, fine-tracking was performed with a broadband visible beacon at Alice's location to compensate for pointing drifts. For this, the QATS employed a fast steering mirror working in the closed-loop mode with a correction for spatial separation between the transmitter main aperture and the beacon.

\begin{figure*}
    \centering
    \includegraphics[width=0.8\linewidth]{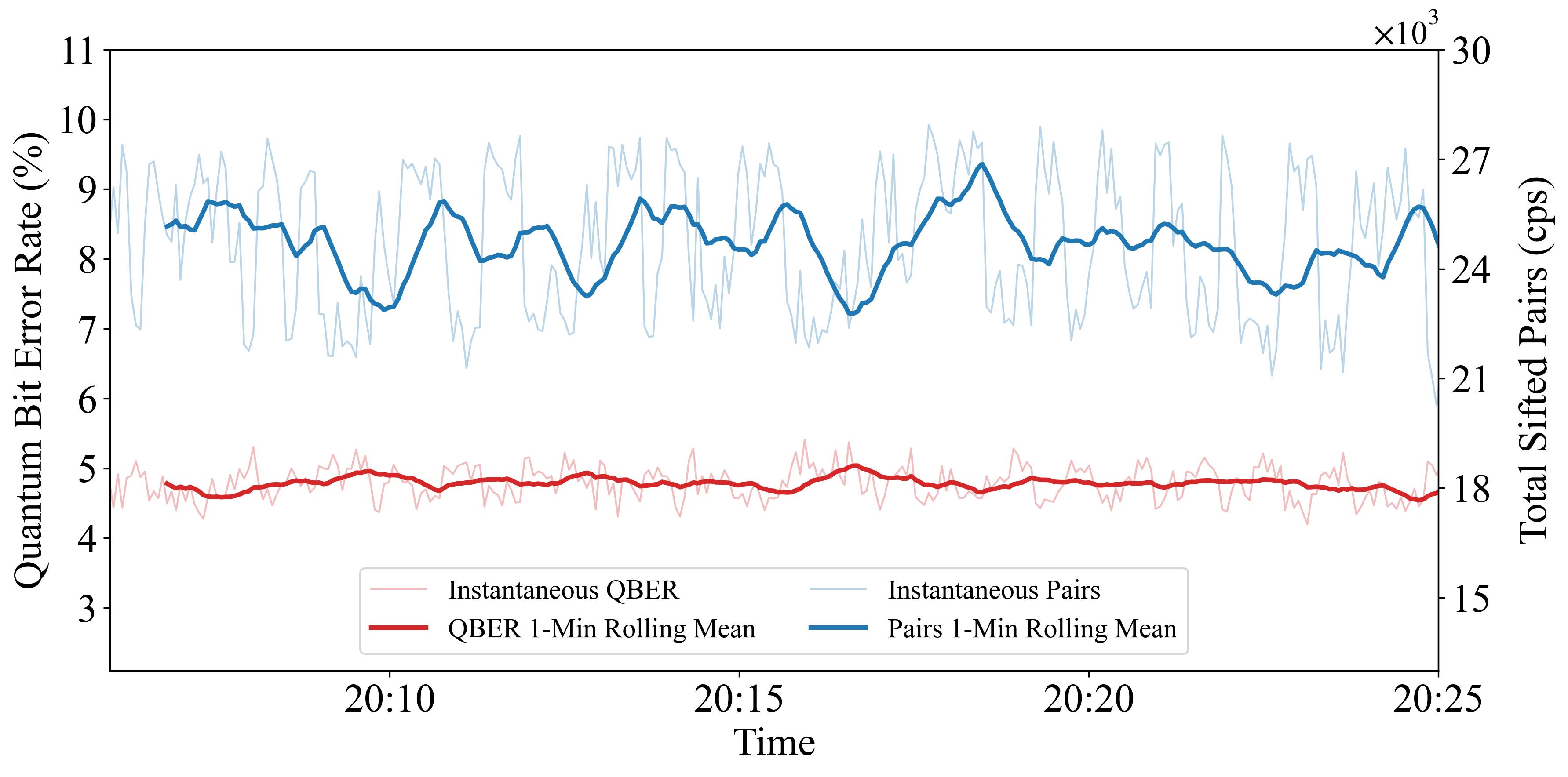}
    \caption{Temporal evolution of system performance metrics. The left axis illustrates the quantum bit error rate, with the instantaneous 1-second resolution data plotted in light red and its corresponding 1-minute rolling mean in dark red. The right axis denotes the detected sifted pair rate in counts per second, scaled by $10^3$. The instantaneous sifted pair rate is depicted in light blue and its 1-minute rolling mean in dark blue. The system's performance remained stable over the whole measurement session.}
    \label{fig:stability}
\end{figure*}

The data acquisition lasted 20~minutes, during which the system exhibited consistent performance without significant degradation in coincidence rates or QBER (Fig. \ref{fig:stability}). A sifted mean coincidence rate of $24,665 \pm 1976$ counts per second (cps) was observed using a coincidence window of $1~\mathrm{ns}$, with a mean QBER of $4.78\,\% \pm 0.24 \, \%$ over the measurement interval. The ambient temperature and relative humidity were $23.6\, ^ \circ$C and $70.7\, \%$, respectively.

\begin{figure}
    \centering
    \includegraphics[width=0.99\linewidth]{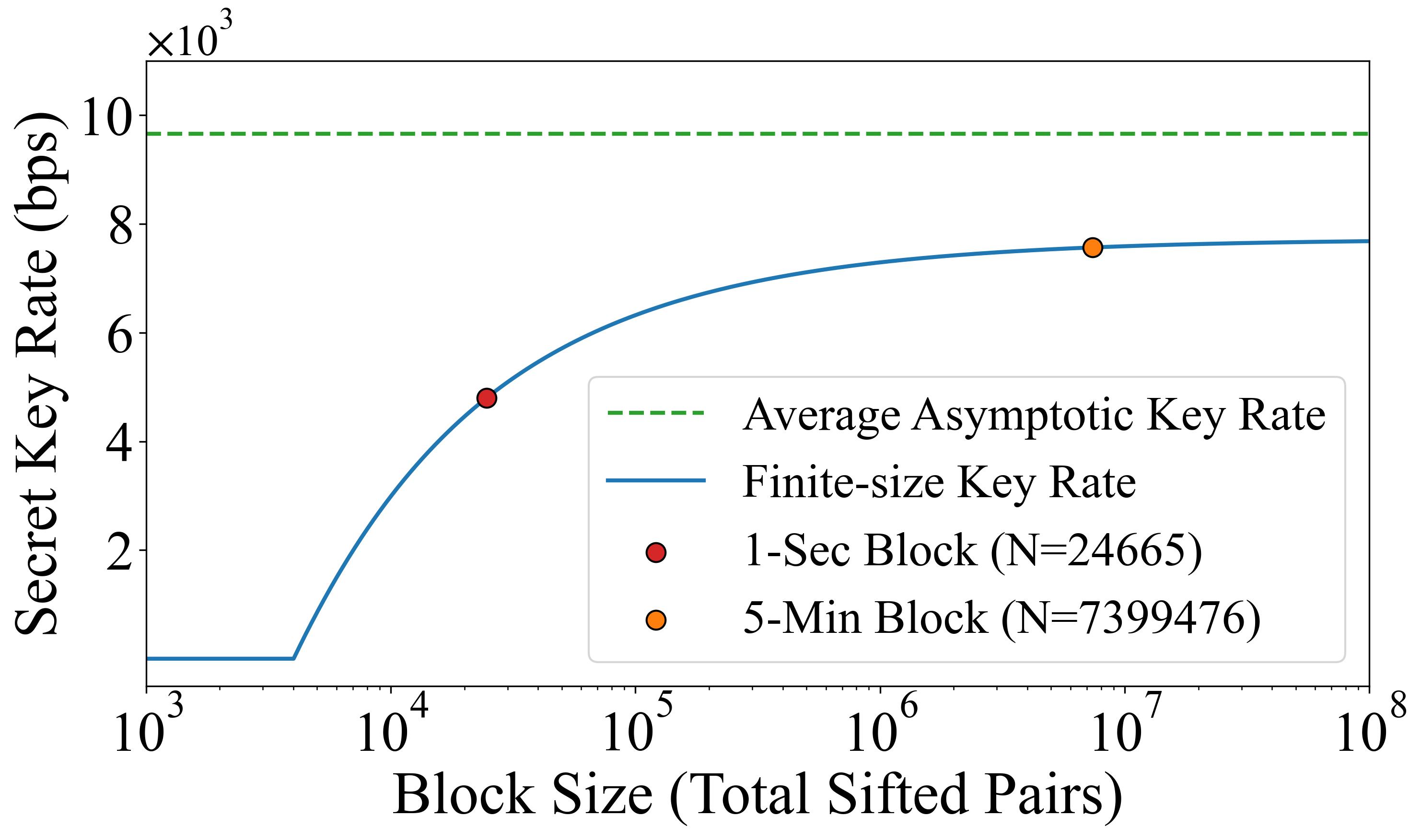}
    \caption{Finite-size secret key rate evaluated as a function of the aggregated block size $N$, using the sharp finite-key bound. The solid blue curve shows the theoretical finite-key prediction, illustrating the impact of statistical finite-size effects for small datasets. The red marker corresponds to 1-second acquisition blocks, while the orange marker represents 5-minute aggregation.}
    \label{fig:keyrate}
\end{figure}

Key sifting was performed using symmetric basis selection at both Alice and Bob, retaining only coincidence events recorded in matching measurement bases for final key generation. All key rates were evaluated on sifted pairs, i.e., after basis reconciliation. For the bit-reconciliation stage we assumed a Low-Density Parity-Check (LDPC) forward-error-correction scheme. Although more efficient error correction protocols (such as Cascade) could in principle yield higher secret key rates, LDPC was selected due to its suitability for forthcoming satellite-to-ground QKD implementations. As a forward error correction scheme, LDPC minimizes bidirectional communication rounds, thereby mitigating the severe throughput penalties typical of high-latency orbital links. To assess the ultimate performance limit of the system, we first evaluated the asymptotic secret key rate following \cite{ma2007quantum} (Fig.~\ref{fig:keyrate}); this represents the theoretical upper bound in the limit of infinite key length. We then evaluated the secret key rate in the finite-size regime. A fixed 20\% of the sifted events was allocated for QBER parameter estimation. While optimizing this sampling fraction could further improve performance, it was kept constant as this optimization falls beyond the scope of the present work. 
The finite-key analysis was carried out within a rigorously composable security framework, providing a protocol-level guarantee conditioned on trusted, ideal detectors, with statistical fluctuations bounded using the recently introduced tight analytical method of Ref.~\cite{mannalath2025sharp}.

To mitigate finite-size effects, the raw key was aggregated over extended acquisition intervals, increasing the effective block size, thereby reducing statistical fluctuations and improving the achievable key rate, at the cost of increased key-generation latency~\cite{islam2024finite}. Aggregating the data into 5-minute blocks yielded a finite-size mean SKR of 7.57~kbps (Fig.~\ref{fig:keyrate}). This corresponds to approximately $80\%$ of the asymptotic rate; the residual gap arises because the parameter-estimation fraction is held fixed at $20\%$ rather than optimized. All calculations used a fixed error-correction efficiency $f = 1.2$, with classical post-processing performed offline after data acquisition. Further implementation details are provided in the Supplementary Material. These results demonstrate efficient entanglement-based QKD operation over a 1.8~km horizontal free-space link under nighttime atmospheric conditions.

\section{Discussion}

To provide a clear development roadmap for future satellite-to-ground missions, we categorize our results into a systematic validation framework for OGS readiness which are summarized in Table \ref{tbl: ogsvalidation}. Our experiment successfully demonstrated several critical `static' operational requirements: the quantum source and receiver optical compatibility, closed-loop pointing of the receiver telescope (Bob) via a classical beacon, and synchronized time-tagging with stable post-processing pipelines. Furthermore, we verified the stability of static polarization basis alignment in an online mode, ensuring that the payload’s internal reference frames remain consistent over the link duration. Notably, our post-processing and synchronization operated without a persistent real-time classical link, though real-time classical communication was still required for the initial polarization alignment.

\begin{table}[t]
\centering
\begin{tabular}{ l c }
\hline
\begin{tabular}{@{}l@{}}\textbf{Validation} \\ \textbf{Metric}\end{tabular} & \begin{tabular}{@{}c@{}}\textbf{Tested during} \\ \textbf{this campaign}\end{tabular} \\ \hline
Quantum signal compatibility & \checkmark \\ \hline
Uplink beacon compatibility & \textsf{X}   \\ \hline
Static pointing and acquisition & \checkmark   \\ \hline
Static synchronization & \checkmark   \\ \hline
Closed-loop tracking & \checkmark\\ \hline
TLE-based tracking & \textsf{X}   \\ \hline
Doppler shift compensation & \textsf{X}   \\ \hline
Post-processing compatibility & \checkmark  \\ \hline
Real-time classical communication & $\textsf{X}^*$   \\ \hline
Static polarization bases alignment & \checkmark \\ \hline
Real-time polarization tracking & \textsf{X}   \\ \hline
Dynamic temporal filtering system & $\textsf{X}^*$   \\ \hline
Key generation at reduced loss level & \checkmark \\ \hline
Key generation at the expected loss level & \textsf{X}   \\ \hline
\end{tabular}
\caption{Summary of OGS operational requirements for a LEO satellite QKD link. (*) Element not required for this specific satellite mission [49].}
\label{tbl: ogsvalidation}
\end{table}

However, several `dynamic' operational requirements remain to be validated to bridge the gap between this 1.8~km terrestrial link and a functional LEO-to-ground link. First, while Bob’s tracking was automated, Alice’s telescope pointing was performed manually in our test. In an operational orbital scenario, the payload must actively lock onto the ADQOGS uplink beacon to achieve the requisite fine-tracking precision. Second, although the OGS is capable of TLE-based (Two-Line Element) acquisition, this was not utilized in the current static experiment. A necessary step is the verification of autonomous orbital tracking based on TLE and dynamic Doppler shift compensation. Third, the relative motion of a satellite pass induces a continuous rotation of the polarization reference frame relative to the ground station. While we validated the internal stability of the payload’s polarization measurement bases, future iterations must implement real-time polarization tracking and correction to maintain high visibility as the satellite's orientation changes throughout the pass. Finally, communication with satellites utilizing pulsed architectures would require an additional dynamic temporal filtering system to mitigate the time-of-flight jitter.

In addition to the `dynamic' requirements, satellite-to-ground links face significant additional losses compared to terrestrial free-space links. The main contribution of loss results from the diffraction of the downlink beam. This is especially prominent in CubeSats such as SpeQtre, where the size constraint limits the aperture of the telescope onboard. 

Thus, while our 1.8~km link validated hardware logic, it operated at a reduced loss level compared to the expected LEO regime. Consequently, key generation at the high-loss regime (25–33~dB) remains a primary objective for future field tests. Moving from static terrestrial verification to the high-velocity, high-loss requirements of an orbital pass provides a benchmark for further qualifying the ground segment readiness for the SpeQtre mission and future satellite QKD operations.

To model the expected key rates in these high-loss regimes, we extrapolate our results to a LEO link. SpeQtre's simulated passes over ADQOGS, which were extrapolated from the satellite's on-orbit TLE obtained from Celestrak \cite{celestrak_tle}, had maximum elevation angles ranging from 25$^\circ$ to 81$^\circ$. Considering the link to be established only when the elevation angle was at least 20$^\circ$, these passes lasted between 162~s to 295~s. For each pass, losses due to beam diffraction and atmospheric absorption and scattering by gases and aerosols were calculated as time-dependent quantities. The QBER, which varies with the signal-to-noise ratio throughout the pass, was similarly calculated at each time step of the extrapolation. 

Assuming an additional 6~dB of loss from tracking errors and turbulence \cite{liao2017satellite, Lu2022}, the total number of sifted coincidences obtained for the pass varied from 2613~counts to 17318~counts. Following the same finite-key analysis, the maximum SKR achieved was 6.7~bps (for the pass with a maximum elevation of 81$^\circ$). These extrapolated results highlight an additional challenge of satellite QKD: the SKR and the total length of secret key obtained can differ greatly from pass to pass.

Looking beyond the immediate requirements for orbital deployment, several advanced technological frontiers can be pursued to enhance the performance and utility of the OGS infrastructure. A primary objective is the integration of adaptive optics (AO) systems at the receiver. By actively mitigating turbulence-induced wavefront distortions, AO offers a pathway to stabilize and optimize coupling efficiency into single-mode fibers \cite{marulanda2024analysis,acosta2024increasing}, a critical requirement for interfacing free-space links with fiber-based quantum repeaters and terrestrial metropolitan networks.
Furthermore, extending the operational window of the station to daylight hours remains a pivotal goal for maximizing the mission duty cycle. This transition will require the implementation of ultra-narrow spectral filtering and the optimization of temporal coincidence windows to maintain high signal-to-noise ratios against solar background radiance. Finally, future iterations could explore the scalability of this architecture for multi-node quantum networks, where the OGS acts as a flexible relay for entanglement swapping or multi-satellite handover protocols \cite{de2025parallel}. These advancements will transition the current architecture from a point-to-point demonstrator into a robust, high-bandwidth node for a global quantum internet.

We also note that the analysis presented in this work follows the standard assumptions adopted for BBM92 implementations \cite{Lim2021}. A dedicated investigation of detector squashing models \cite{Tsurumaru2008squash} and the effect of detector-side imperfections \cite{Xu2020realisticDevices} on the secret key rate is left for future work.

In summary, we have successfully demonstrated the operational integration of a space-qualified quantum payload with an automated optical ground station. The achieved key rates and stability metrics verify the hardware compatibility and background rejection strategies required for space-to-ground quantum communication. These results confirm the terrestrial validation of the architecture for the SpeQtre mission, marking a decisive step toward the deployment of reliable satellite-based quantum networks.

\section*{Abbreviations}
\begin{itemize}
    \item \textbf{ADQOGS:} Abu Dhabi Quantum Optical Ground Station
    \item \textbf{OGS:} Optical Ground Station
    \item \textbf{AO:} Adaptive Optics
    \item \textbf{AOD:} Aerosol Optical Depth
    \item \textbf{EB:} Entanglement-based
    \item \textbf{FWHM:} Full width at half maximum
    \item \textbf{GPS:} Global Positioning System
    \item \textbf{LDPC:} Low-Density Parity-Check
    \item \textbf{LEO:} Low Earth Orbit
    \item \textbf{PAT:} Pointing, Acquisition, and Tracking
    \item \textbf{QATS:} Quantum Acquisition and Tracking System
    \item \textbf{QBER:} Quantum Bit Error Rate
    \item \textbf{QKD:} Quantum Key Distribution
    \item \textbf{QM:} Quantum Module
    \item \textbf{SKR:} Secret Key Rate
    \item \textbf{SPD:} Single-Photon Detector
    \item \textbf{SPDC:} Spontaneous Parametric Down Conversion
    \item \textbf{TLE:} Two-Line Element
    
\end{itemize}

\section*{Declarations}
\begin{itemize}
\item \textbf{Availability of data and materials:} The datasets used and during the current study are available from the corresponding author upon reasonable request.
\item \textbf{Competing interests:} The authors declare that they have no competing interests
\item \textbf{Funding:} 
This project is supported by the National Research Foundation, Singapore through the National Quantum Office, hosted in A*STAR, under its Centre for Quantum Technologies Funding Initiative (S24Q2d0009).

\item \textbf{Authors' contributions:} G.D.S., J.B.C., S.S., K.K., and C.C.L. executed the field experiments. G.D.S., J.B.C., and A.V. were responsible for data analysis and preparation of the initial manuscript draft. All authors provided critical feedback, reviewed and approved the final manuscript.
\item \textbf{Acknowledgements:} We thank Fabian Koehler for valuable initial discussions regarding the finite-key analysis, and Abdurahoof Edakkanden for his assistance with the deployment of the experiment. We also thank the Pawan Gupta for his effort in establishing and maintaining Masdar Institute 2 as part of the Aerosol Robotic Network (AERONET).

\end{itemize}

\bibliography{biblio.bib}

@article{wei2022towards,
  title={Towards real-world quantum networks: a review},
  author={Wei, Shi-Hai and Jing, Bo and Zhang, Xue-Ying and Liao, Jin-Yu and Yuan, Chen-Zhi and Fan, Bo-Yu and Lyu, Chen and Zhou, Dian-Li and Wang, You and Deng, Guang-Wei and others},
  journal={Laser \& Photonics Reviews},
  volume={16},
  number={3},
  pages={2100219},
  year={2022},
  publisher={Wiley Online Library},
  URL={https://doi.org/10.1002/lpor.202100219}
  
}

@article{kimble2008quantum,
  title={The quantum internet},
  author={Kimble, H Jeff},
  journal={Nature},
  volume={453},
  number={7198},
  pages={1023--1030},
  year={2008},
  publisher={Nature Publishing Group},
  URL={https://doi.org/10.1038/nature07127}
}

@article{wehner2018quantum,
  title={Quantum internet: A vision for the road ahead},
  author={Wehner, Stephanie and Elkouss, David and Hanson, Ronald},
  journal={Science},
  volume={362},
  number={6412},
  pages={eaam9288},
  year={2018},
  publisher={American Association for the Advancement of Science},
  URL={https://doi.org/10.1126/science.aam9288}
}

@article{simon2017towards,
  title={Towards a global quantum network},
  author={Simon, Christoph},
  journal={Nature Photonics},
  volume={11},
  number={11},
  pages={678--680},
  year={2017},
  publisher={Nature Publishing Group UK London},
  URL={https://doi.org/10.1038/s41566-017-0032-0}
}

@article{de2023satellite,
  title={Satellite-based quantum information networks: use cases, architecture, and roadmap},
  author={de Forges de Parny, Laurent and Alibart, Olivier and Debaud, Julien and Gressani, Sacha and Lagarrigue, Alek and Martin, Anthony and Metrat, Alexandre and Schiavon, Matteo and Troisi, Tess and Diamanti, Eleni and others},
  journal={Communications Physics},
  volume={6},
  number={1},
  pages={12},
  year={2023},
  publisher={Nature Publishing Group UK London},
  URL={https://doi.org/10.1038/s42005-022-01123-7}
}

@article{gisin2002quantum,
  title={Quantum cryptography},
  author={Gisin, Nicolas and Ribordy, Gr{\'e}goire and Tittel, Wolfgang and Zbinden, Hugo},
  journal={Reviews of modern physics},
  volume={74},
  number={1},
  pages={145},
  year={2002},
  publisher={APS},
  URL={https://doi.org/10.1103/RevModPhys.74.145}
}

@article{acin2006bell,
  title={From Bell’s theorem to secure quantum key distribution},
  author={Acin, Antonio and Gisin, Nicolas and Masanes, Lluis},
  journal={Physical review letters},
  volume={97},
  number={12},
  pages={120405},
  year={2006},
  publisher={APS},
  URL={https://doi.org/10.1103/PhysRevLett.97.120405}
}

@article{zapatero2023advances,
  title={Advances in device-independent quantum key distribution},
  author={Zapatero, V{\'\i}ctor and van Leent, Tim and Arnon-Friedman, Rotem and Liu, Wen-Zhao and Zhang, Qiang and Weinfurter, Harald and Curty, Marcos},
  journal={npj quantum information},
  volume={9},
  number={1},
  pages={10},
  year={2023},
  publisher={Nature Publishing Group UK London},
  URL={https://doi.org/10.1038/s41534-023-00684-x}
}

@article{pan1998experimental,
  title={Experimental entanglement swapping: entangling photons that never interacted},
  author={Pan, Jian-Wei and Bouwmeester, Dik and Weinfurter, Harald and Zeilinger, Anton},
  journal={Physical review letters},
  volume={80},
  number={18},
  pages={3891},
  year={1998},
  publisher={APS},
  URL={https://doi.org/10.1103/PhysRevLett.80.3891}
}

@article{azuma2023quantum,
  title={Quantum repeaters: From quantum networks to the quantum internet},
  author={Azuma, Koji and Economou, Sophia E and Elkouss, David and Hilaire, Paul and Jiang, Liang and Lo, Hoi-Kwong and Tzitrin, Ilan},
  journal={Reviews of Modern Physics},
  volume={95},
  number={4},
  pages={045006},
  year={2023},
  publisher={APS},
  URL={https://doi.org/10.1103/RevModPhys.95.045006}
}

@inproceedings{broadbent2009universal,
  title={Universal blind quantum computation},
  author={Broadbent, Anne and Fitzsimons, Joseph and Kashefi, Elham},
  booktitle={2009 50th annual IEEE symposium on foundations of computer science},
  pages={517--526},
  year={2009},
  organization={IEEE},
  URL={https://doi.org/10.1109/FOCS.2009.36}
}

@inproceedings{yimsiriwattana2004distributed,
  title={Distributed quantum computing: A distributed Shor algorithm},
  author={Yimsiriwattana, Anocha and Lomonaco Jr, Samuel J},
  booktitle={Quantum information and computation II},
  volume={5436},
  pages={360--372},
  year={2004},
  organization={SPIE},
  URL={https://doi.org/10.1117/12.546504}
}

@article{zhang2021distributed,
  title={Distributed quantum sensing},
  author={Zhang, Zheshen and Zhuang, Quntao},
  journal={Quantum Science and Technology},
  volume={6},
  number={4},
  pages={043001},
  year={2021},
  publisher={IOP Publishing},
  URL={https://doi.org/10.1088/2058-9565/abd4c3}
}

@article{pirandola2017fundamental,
  title={Fundamental limits of repeaterless quantum communications},
  author={Pirandola, Stefano and Laurenza, Riccardo and Ottaviani, Carlo and Banchi, Leonardo},
  journal={Nature communications},
  volume={8},
  number={1},
  pages={15043},
  year={2017},
  publisher={Nature Publishing Group UK London},
  URL={https://doi.org/10.1038/ncomms15043}
}

@article{yin2016measurement,
  title={Measurement-device-independent quantum key distribution over a 404 km optical fiber},
  author={Yin, Hua-Lei and Chen, Teng-Yun and Yu, Zong-Wen and Liu, Hui and You, Li-Xing and Zhou, Yi-Heng and Chen, Si-Jing and Mao, Yingqiu and Huang, Ming-Qi and Zhang, Wei-Jun and others},
  journal={Physical review letters},
  volume={117},
  number={19},
  pages={190501},
  year={2016},
  publisher={APS},
  URL={https://doi.org/10.1103/PhysRevLett.117.190501}
}

@article{liu2023experimental,
  title={Experimental twin-field quantum key distribution over 1000 km fiber distance},
  author={Liu, Yang and Zhang, Wei-Jun and Jiang, Cong and Chen, Jiu-Peng and Zhang, Chi and Pan, Wen-Xin and Ma, Di and Dong, Hao and Xiong, Jia-Min and Zhang, Cheng-Jun and others},
  journal={Physical Review Letters},
  volume={130},
  number={21},
  pages={210801},
  year={2023},
  publisher={APS},
  URL={https://doi.org/10.1103/PhysRevLett.130.210801}
}

@article{li2023high,
  title={High-rate quantum key distribution exceeding 110 Mb s--1},
  author={Li, Wei and Zhang, Likang and Tan, Hao and Lu, Yichen and Liao, Sheng-Kai and Huang, Jia and Li, Hao and Wang, Zhen and Mao, Hao-Kun and Yan, Bingze and others},
  journal={Nature photonics},
  volume={17},
  number={5},
  pages={416--421},
  year={2023},
  publisher={Nature Publishing Group UK London},
  URL={https://doi.org/10.1038/s41566-023-01166-4}
}

@article{yuan201810,
  title={10-Mb/s quantum key distribution},
  author={Yuan, Zhiliang and Plews, Alan and Takahashi, Ririka and Doi, Kazuaki and Tam, Winci and Sharpe, Andrew and Dixon, Alexander and Lavelle, Evan and Dynes, James and Murakami, Akira and others},
  journal={Journal of Lightwave Technology},
  volume={36},
  number={16},
  pages={3427--3433},
  year={2018},
  publisher={OSA},
  URL={https://doi.org/10.1109/JLT.2018.2843136}
}

@article{frohlich2015quantum,
  title={Quantum secured gigabit optical access networks},
  author={Fr{\"o}hlich, Bernd and Dynes, James F and Lucamarini, Marco and Sharpe, Andrew W and Tam, Simon W-B and Yuan, Zhiliang and Shields, Andrew J},
  journal={Scientific reports},
  volume={5},
  number={1},
  pages={18121},
  year={2015},
  publisher={Nature Publishing Group UK London},
  URL={https://doi.org/10.1038/srep18121}
}

@techreport{oecd_rights_of_way_fibre_2008,
  author      = {{OECD}},
  title       = {Public Rights of Way for Fibre Deployment to the Home},
  year        = {2008},
  institution = {OECD},
  url         = {https://www.oecd.org/content/dam/oecd/en/publications/reports/2008/07/public-rights-of-way-for-fibre-deployment-to-the-home_g17a1c1a/230502835656.pdf},
  note        = {Accessed 2026-01-13}
}

@misc{au_crossborder_fibre_permits,
  author       = {{African Union}},
  title        = {Cross Border and Interconnection Policies},
  howpublished = {PDF},
  url          = {https://au.int/sites/default/files/documents/31043-doc-03-cross-border-and-interconnection-policies-en.pdf},
  note         = {Accessed 2026-01-13}
}

@inproceedings{amairi2024versatile,
  title={Versatile optical ground station for satellite-based quantum key distribution in Abu Dhabi},
  author={Amairi-Pyka, Sana and Fischer, Christoph and Kravtsov, Konstantin and De Santis, Gianluca and Grosso, Alessandro and Fischer, Edgar and Kudielka, Klaus and Grieve, James A},
  volume = {13699},
  booktitle = {International Conference on Space Optics — ICSO 2024},
  organization = {International Society for Optics and Photonics},
  publisher = {SPIE},
  pages = {136993N},
  year = {2025},
  URL={https://doi.org/10.1117/12.3075235}
}

@article{yin2020entanglement,
  title={Entanglement-based secure quantum cryptography over 1,120 kilometres},
  author={Yin, Juan and Li, Yu-Huai and Liao, Sheng-Kai and Yang, Meng and Cao, Yuan and Zhang, Liang and Ren, Ji-Gang and Cai, Wen-Qi and Liu, Wei-Yue and Li, Shuang-Lin and others},
  journal={Nature},
  volume={582},
  number={7813},
  pages={501--505},
  year={2020},
  publisher={Nature Publishing Group UK London}
}

@article{liao2018satellite,
  title={Satellite-relayed intercontinental quantum network},
  author={Liao, Sheng-Kai and Cai, Wen-Qi and Handsteiner, Johannes and Liu, Bo and Yin, Juan and Zhang, Liang and Rauch, Dominik and Fink, Matthias and Ren, Ji-Gang and Liu, Wei-Yue and others},
  journal={Physical review letters},
  volume={120},
  number={3},
  pages={030501},
  year={2018},
  publisher={APS},
  URL={https://doi.org/10.1103/PhysRevLett.120.030501}
}

@article{liao2017satellite,
  title={Satellite-to-ground quantum key distribution},
  author={Liao, Sheng-Kai and Cai, Wen-Qi and Liu, Wei-Yue and Zhang, Liang and Li, Yang and Ren, Ji-Gang and Yin, Juan and Shen, Qi and Cao, Yuan and Li, Zheng-Ping and others},
  journal={Nature},
  volume={549},
  number={7670},
  pages={43--47},
  year={2017},
  publisher={Nature Publishing Group UK London},
  URL={https://doi.org/10.1038/nature23655}
}

@article{li2025microsatellite,
  title={Microsatellite-based real-time quantum key distribution},
  author={Li, Yang and Cai, Wen-Qi and Ren, Ji-Gang and Wang, Chao-Ze and Yang, Meng and Zhang, Liang and Wu, Hui-Ying and Chang, Liang and Wu, Jin-Cai and Jin, Biao and others},
  journal={Nature},
  pages={1--8},
  year={2025},
  publisher={Nature Publishing Group UK London},
  URL={https://doi.org/10.1038/s41586-025-08739-z}
}

@article{liao2017space,
  title={Space-to-ground quantum key distribution using a small-sized payload on Tiangong-2 space lab},
  author={Liao, Sheng-Kai and Lin, Jin and Ren, Ji-Gang and Liu, Wei-Yue and Qiang, Jia and Yin, Juan and Li, Yang and Shen, Qi and Zhang, Liang and Liang, Xue-Feng and others},
  journal={Chinese Physics Letters},
  volume={34},
  number={9},
  pages={090302},
  year={2017},
  publisher={IOP Publishing},
  URL={https://doi.org/10.1088/0256-307X/34/9/090302}
}

@article{khmelev2024eurasian,
  title={Eurasian-scale experimental satellite-based quantum key distribution with detector efficiency mismatch analysis},
  author={Khmelev, Aleksandr and Duplinsky, Alexey and Bakhshaliev, Ruslan and Ivchenko, Egor and Pismeniuk, Liubov and Mayboroda, Vladimir and Nesterov, Ivan and Chernov, Arkadiy and Trushechkin, Anton and Kiktenko, Evgeniy and others},
  journal={Optics Express},
  volume={32},
  number={7},
  pages={11964--11978},
  year={2024},
  publisher={Optica Publishing Group},
  URL={https://doi.org/10.1364/OE.511772}
}

@article{mishra2022bbm92,
  title={BBM92 quantum key distribution over a free space dusty channel of 200 meters},
  author={Mishra, Sarika and Biswas, Ayan and Patil, Satyajeet and Chandravanshi, Pooja and Mongia, Vardaan and Sharma, Tanya and Rani, Anju and Prabhakar, Shashi and Ramachandran, S and Singh, Ravindra P},
  journal={Journal of Optics},
  volume={24},
  number={7},
  pages={074002},
  year={2022},
  publisher={IOP Publishing},
  URL={https://doi.org/10.1088/2040-8986/ac6f0b}
}

@article{cao2013entanglement,
  title={Entanglement-based quantum key distribution with biased basis choice via free space},
  author={Cao, Yuan and Liang, Hao and Yin, Juan and Yong, Hai-Lin and Zhou, Fei and Wu, Yu-Ping and Ren, Ji-Gang and Li, Yu-Huai and Pan, Ge-Sheng and Yang, Tao and others},
  journal={Optics express},
  volume={21},
  number={22},
  pages={27260--27268},
  year={2013},
  publisher={Optical Society of America},
  URL={https://doi.org/10.1364/OE.21.027260}
}

@article{erven2012studying,
  title={Studying free-space transmission statistics and improving free-space quantum key distribution in the turbulent atmosphere},
  author={Erven, Christopher and Heim, B and Meyer-Scott, E and Bourgoin, JP and Laflamme, R and Weihs, G and Jennewein, T},
  journal={New Journal of Physics},
  volume={14},
  number={12},
  pages={123018},
  year={2012},
  publisher={IOP Publishing},
  URL={https://doi.org/10.1088/1367-2630/14/12/123018}
}

@article{li2022space,
  title={Space--ground QKD network based on a compact payload and medium-inclination orbit},
  author={Li, Yang and Liao, Sheng-Kai and Cao, Yuan and Ren, Ji-Gang and Liu, Wei-Yue and Yin, Juan and Shen, Qi and Qiang, Jia and Zhang, Liang and Yong, Hai-Lin and others},
  journal={Optica},
  volume={9},
  number={8},
  pages={933--938},
  year={2022},
  publisher={Optica Publishing Group},
  URL={https://doi.org/10.1364/OPTICA.458330}
}

@article{yin2017satellite,
  title={Satellite-based entanglement distribution over 1200 kilometers},
  author={Yin, Juan and Cao, Yuan and Li, Yu-Huai and Liao, Sheng-Kai and Zhang, Liang and Ren, Ji-Gang and Cai, Wen-Qi and Liu, Wei-Yue and Li, Bo and Dai, Hui and others},
  journal={Science},
  volume={356},
  number={6343},
  pages={1140--1144},
  year={2017},
  publisher={American Association for the Advancement of Science},
  URL={https://doi.org/10.1126/science.aan3211}
}

@article{villar2020entanglement,
  title={Entanglement demonstration on board a nano-satellite},
  author={Villar, Aitor and Lohrmann, Alexander and Bai, Xueliang and Vergoossen, Tom and Bedington, Robert and Perumangatt, Chithrabhanu and Lim, Huai Ying and Islam, Tanvirul and Reezwana, Ayesha and Tang, Zhongkan and others},
  journal={Optica},
  volume={7},
  number={7},
  pages={734--737},
  year={2020},
  publisher={Optical Society of America}
}

@article{yin2017satellite_b,
  title={Satellite-to-ground entanglement-based quantum key distribution},
  author={Yin, Juan and Cao, Yuan and Li, Yu-Huai and Ren, Ji-Gang and Liao, Sheng-Kai and Zhang, Liang and Cai, Wen-Qi and Liu, Wei-Yue and Li, Bo and Dai, Hui and others},
  journal={Physical review letters},
  volume={119},
  number={20},
  pages={200501},
  year={2017},
  publisher={APS},
  URL={https://doi.org/10.1103/PhysRevLett.119.200501}
}

@article{sidhu2021advances,
  title={Advances in space quantum communications},
  author={Sidhu, Jasminder S and Joshi, Siddarth K and G{\"u}ndo{\u{g}}an, Mustafa and Brougham, Thomas and Lowndes, David and Mazzarella, Luca and Krutzik, Markus and Mohapatra, Sonali and Dequal, Daniele and Vallone, Giuseppe and others},
  journal={IET Quantum Communication},
  volume={2},
  number={4},
  pages={182--217},
  year={2021},
  publisher={Wiley Online Library},
  URL={https://doi.org/10.1049/qtc2.12015}
}

@article{bedington2017progress,
  title={Progress in satellite quantum key distribution},
  author={Bedington, Robert and Arrazola, Juan Miguel and Ling, Alexander},
  journal={npj Quantum Information},
  volume={3},
  number={1},
  pages={30},
  year={2017},
  publisher={Nature Publishing Group UK London},
  URL={https://doi.org/10.1038/s41534-017-0031-5}
}

@article{peng2005experimental,
  title={Experimental Free-Space Distribution of Entangled Photon Pairs Over 13 km:<? format?> Towards Satellite-Based Global Quantum Communication},
  author={Peng, Cheng-Zhi and Yang, Tao and Bao, Xiao-Hui and Zhang, Jun and Jin, Xian-Min and Feng, Fa-Yong and Yang, Bin and Yang, Jian and Yin, Juan and Zhang, Qiang and others},
  journal={Physical review letters},
  volume={94},
  number={15},
  pages={150501},
  year={2005},
  publisher={APS},
  URL={https://doi.org/10.1103/PhysRevLett.94.150501}
}

@article{marcikic2006free,
  title={Free-space quantum key distribution with entangled photons},
  author={Marcikic, Ivan and Lamas-Linares, Antia and Kurtsiefer, Christian},
  journal={Applied Physics Letters},
  volume={89},
  number={10},
  year={2006},
  publisher={AIP Publishing},
  URL={https://doi.org/10.1063/1.2348775}
}

@article{ursin2007entanglement,
  title={Entanglement-based quantum communication over 144 km},
  author={Ursin, Rupert and Tiefenbacher, Felix and Schmitt-Manderbach, Tobias and Weier, Henning and Scheidl, Thomas and Lindenthal, Michael and Blauensteiner, Bibiane and Jennewein, Thomas and Perdigues, Josep and Trojek, Pavel and others},
  journal={Nature physics},
  volume={3},
  number={7},
  pages={481--486},
  year={2007},
  publisher={Nature Publishing Group UK London},
  URL={https://doi.org/10.1038/nphys629}
}

@article{erven2008entangled,
  title={Entangled quantum key distribution over two free-space optical links},
  author={Erven, Christopher and Couteau, C and Laflamme, R and Weihs, G},
  journal={Optics express},
  volume={16},
  number={21},
  pages={16840--16853},
  year={2008},
  publisher={Optical Society of America},
  URL={https://doi.org/10.1364/OE.16.016840}
}

@article{ling2008experimental,
  title={Experimental quantum key distribution based on a Bell test},
  author={Ling, Alexander and Peloso, Matthew P and Marcikic, Ivan and Scarani, Valerio and Lamas-Linares, Antia and Kurtsiefer, Christian},
  journal={Physical Review A—Atomic, Molecular, and Optical Physics},
  volume={78},
  number={2},
  pages={020301},
  year={2008},
  publisher={APS},
  URL={https://doi.org/10.1103/PhysRevA.78.020301}
}

@article{peloso2009daylight,
  title={Daylight operation of a free space, entanglement-based quantum key distribution system},
  author={Peloso, Matthew P and Gerhardt, Ilja and Ho, Caleb and Lamas-Linares, Antia and Kurtsiefer, Christian},
  journal={New Journal of Physics},
  volume={11},
  number={4},
  pages={045007},
  year={2009},
  publisher={IOP Publishing},
  URL={https://doi.org/10.1088/1367-2630/11/4/045007}
}

@article{scheidl2009feasibility,
  title={Feasibility of 300 km quantum key distribution with entangled states},
  author={Scheidl, Thomas and Ursin, Rupert and Fedrizzi, Alessandro and Ramelow, Sven and Ma, Xiao-Song and Herbst, Thomas and Prevedel, Robert and Ratschbacher, Lothar and Kofler, Johannes and Jennewein, Thomas and others},
  journal={New Journal of Physics},
  volume={11},
  number={8},
  pages={085002},
  year={2009},
  publisher={IOP Publishing},
  URL={https://doi.org/10.1088/1367-2630/11/8/085002}
}

@article{ecker2021strategies,
  title={Strategies for achieving high key rates in satellite-based QKD},
  author={Ecker, Sebastian and Liu, Bo and Handsteiner, Johannes and Fink, Matthias and Rauch, Dominik and Steinlechner, Fabian and Scheidl, Thomas and Zeilinger, Anton and Ursin, Rupert},
  journal={npj Quantum Information},
  volume={7},
  number={1},
  pages={5},
  year={2021},
  publisher={Nature Publishing Group UK London},
  URL={https://doi.org/10.1038/s41534-020-00335-5}
}

@article{krvzivc2023towards,
  title={Towards metropolitan free-space quantum networks},
  author={Kr{\v{z}}i{\v{c}}, Andrej and Sharma, Sakshi and Spiess, Christopher and Chandrashekara, Uday and T{\"o}pfer, Sebastian and Sauer, Gregor and Gonz{\'a}lez-Mart{\'\i}n del Campo, Luis Javier and Kopf, Teresa and Petscharnig, Stefan and Grafenauer, Thomas and others},
  journal={npj Quantum Information},
  volume={9},
  number={1},
  pages={95},
  year={2023},
  publisher={Nature Publishing Group UK London},
  URL={https://doi.org/10.1038/s41534-023-00754-0}
}

@inproceedings{10.52202/083082-0038,
    author = {De Santis, Gianluca and Kravtsov, Konstantin and Kara, Ozan and Amairi Pyka, Sana and Grieve, James A.},
    title = {Site Characterization for Satellite Quantum Key Distribution at the Abu Dhabi Quantum Optical Ground Station},
    booktitle = {IAF Space Communications and Navigation Symposium},
    publisher = {International Astronautical Federation (IAF)},
    pages = {318-326},
    year = {2025},
    doi = {10.52202/083082-0038},
    URL = {https://doi.org/10.52202/083082-0038}
}

@article{Lohrmann2020,
  author     = {Alexander Lohrmann and Chithrabhanu Perumangatt and Aitor Villar and Alexander Ling},
  journal    = {Appl. Phys. Lett.},
  title      = {Broadband pumped polarization entangled photon-pair source in a linear beam displacement interferometer},
  year       = {2020},
  month      = {jan},
  number     = {2},
  pages      = {021101},
  volume     = {116},
  doi        = {10.1063/1.5124416},
  fjournal   = {Applied Physics Letters},
  publisher  = {{AIP} Publishing}
}

@Article{Lu2022,
  author    = {Lu, Chao-Yang and Cao, Yuan and Peng, Cheng-Zhi and Pan, Jian-Wei},
  journal   = {Reviews of Modern Physics},
  title     = {Micius quantum experiments in space},
  year      = {2022},
  issn      = {1539-0756},
  month     = jul,
  number    = {3},
  pages     = {035001},
  volume    = {94},
  doi       = {10.1103/revmodphys.94.035001},
  publisher = {American Physical Society (APS)},
}

@article{mannalath2025sharp,
  title={Sharp finite statistics for quantum key distribution},
  author={Mannalath, Vaisakh and Zapatero, V{\'\i}ctor and Curty, Marcos},
  journal={Physical Review Letters},
  volume={135},
  number={2},
  pages={020803},
  year={2025},
  publisher={APS}
}

@article{islam2024finite,
  title={Finite-resource performance of small-satellite-based quantum-key-distribution missions},
  author={Islam, Tanvirul and Sidhu, Jasminder S and Higgins, Brendon L and Brougham, Thomas and Vergoossen, Tom and Oi, Daniel KL and Jennewein, Thomas and Ling, Alexander},
  journal={PRX Quantum},
  volume={5},
  number={3},
  pages={030101},
  year={2024},
  publisher={APS}
}

@article{ma2007quantum,
  title={Quantum key distribution with entangled photon sources},
  author={Ma, Xiongfeng and Fung, Chi-Hang Fred and Lo, Hoi-Kwong},
  journal={Physical Review A—Atomic, Molecular, and Optical Physics},
  volume={76},
  number={1},
  pages={012307},
  year={2007},
  publisher={APS}
}

@Misc{Rozenman2026,
  author    = {Rozenman, Georgi Gary and Maslennikov, Alona and Gandelman, Sara P. and Reches, Yuval and Delfan, Sahar and Kundu, Neel Kanth and Zhang, Leyi and Liu, Ruiqi},
  title     = {Free-space and Satellite-Based Quantum Communication: Principles, Implementations, and Challenges},
  year      = {2026},
  copyright = {Creative Commons Attribution Non Commercial No Derivatives 4.0 International},
  doi       = {10.48550/ARXIV.2602.01426},
  publisher = {arXiv},
}

@InProceedings{Trinh2018,
  author    = {Trinh, Phuc V. and Pham, Anh T. and Carrasco-Casado, Alberto and Toyoshima, Moria},
  booktitle = {2018 Progress in Electromagnetics Research Symposium (PIERS-Toyama)},
  title     = {Quantum Key Distribution over FSO: Current Development and Future Perspectives},
  year      = {2018},
  month     = aug,
  pages     = {1672--1679},
  publisher = {IEEE},
  doi       = {10.23919/piers.2018.8597918},
}

@Article{Mehic2020,
  author    = {Mehic, Miralem and Niemiec, Marcin and Rass, Stefan and Ma, Jiajun and Peev, Momtchil and Aguado, Alejandro and Martin, Vicente and Schauer, Stefan and Poppe, Andreas and Pacher, Christoph and Voznak, Miroslav},
  journal   = {ACM Computing Surveys},
  title     = {Quantum Key Distribution: A Networking Perspective},
  year      = {2020},
  issn      = {1557-7341},
  month     = sep,
  number    = {5},
  pages     = {1--41},
  volume    = {53},
  doi       = {10.1145/3402192},
  publisher = {Association for Computing Machinery (ACM)},
}

@article{de2025parallel,
  title={Parallel trusted node approach for satellite quantum key distribution},
  author={De Santis, Gianluca and Kravtsov, Konstantin and Amairi-Pyka, Sana and Grieve, James A},
  journal={EPJ Quantum Technology},
  volume={12},
  number={1},
  pages={50},
  year={2025},
  publisher={Springer}
}

@article{marulanda2024analysis,
  title={Analysis of satellite-to-ground quantum key distribution with adaptive optics},
  author={Marulanda Acosta, Valentina and Dequal, Daniele and Schiavon, Matteo and Montmerle-Bonnefois, Aur{\'e}lie and Lim, Caroline B and Conan, Jean-Marc and Diamanti, Eleni},
  journal={New Journal of Physics},
  volume={26},
  number={2},
  pages={023039},
  year={2024},
  publisher={IOP Publishing}
}

@article{acosta2024increasing,
  title={Increasing the secret key rate of satellite-to-ground entanglement-based QKD assisted by adaptive optics},
  author={Acosta, Valentina Marulanda and Dequal, Daniele and Schiavon, Matteo and Montmerle-Bonnefois, Aurelie and Lim, Caroline B and Conan, Jean-Marc and Diamanti, Eleni},
  journal={arXiv preprint arXiv:2411.09564},
  year={2024}
}

@inproceedings{villar2026speqtre,
  title={SpeQtre: a 12U CubeSat to demonstrate entanglement-based quantum communications from space to ground},
  author={Villar, Aitor},
  booktitle={Quantum Computing, Communication, and Simulation VI},
  pages={PC1391907},
  year={2026},
  organization={SPIE}
}

@Article{Xu2020realisticDevices,
  author    = {Xu, Feihu and Ma, Xiongfeng and Zhang, Qiang and Lo, Hoi-Kwong and Pan, Jian-Wei},
  journal   = {Reviews of Modern Physics},
  title     = {Secure quantum key distribution with realistic devices},
  year      = {2020},
  issn      = {1539-0756},
  month     = may,
  number    = {2},
  pages     = {025002},
  volume    = {92},
  doi       = {10.1103/revmodphys.92.025002},
  publisher = {American Physical Society (APS)},
}

@Article{Tsurumaru2008squash,
  author    = {Tsurumaru, Toyohiro and Tamaki, Kiyoshi},
  journal   = {Physical Review A},
  title     = {Security proof for quantum-key-distribution systems with threshold detectors},
  year      = {2008},
  issn      = {1094-1622},
  month     = Sept,
  number    = {3},
  pages     = {032302},
  volume    = {78},
  doi       = {10.1103/physreva.78.032302},
  publisher = {American Physical Society (APS)},
}

@Article{Lim2021,
  author    = {Lim, Charles Ci-Wen and Xu, Feihu and Pan, Jian-Wei and Ekert, Artur},
  journal   = {Physical Review Letters},
  title     = {Security Analysis of Quantum Key Distribution with Small Block Length and Its Application to Quantum Space Communications},
  year      = {2021},
  issn      = {1079-7114},
  month     = mar,
  number    = {10},
  pages     = {100501},
  volume    = {126},
  doi       = {10.1103/physrevlett.126.100501},
  publisher = {American Physical Society (APS)},
}

@dataset{celestrak_tle,
  author       = {Kelso, T. S.},
  title        = {{CelesTrak Two-Line Element Set for NORAD ID 66769 (Epoch: 2026-07-13 18:49)}},
  year         = {2026},
  url          = {https://celestrak.org},
  note         = {Accessed: 14 July 2026}
}
\end{document}